\documentstyle[epsf,epsfig,rotate]{mn}

\newcommand{\be}{\begin{equation}}
\newcommand{\ee}{\end{equation}}
\newcommand{\ben}{\begin{eqnarray}}
\newcommand{\een}{\end{eqnarray}}

\newcommand{\beq}{\begin{equation}}
\newcommand{\eeq}{\end{equation}}
\newcommand{\bde}{\begin{displaymath}}
\newcommand{\ede}{\end{displaymath}}

\newcommand{\bmin}[1]{\begin{minipage}[b]{#1\linewidth}}
\newcommand{\emin}{\end{minipage}}
\newcommand{\bfig}{\begin{figure}}
\newcommand{\efig}{\end{figure}}

\renewcommand{\sun}{\hbox{$_\odot$}}
\newcommand{\au}     {\hbox{$\,\mathrm{au}$}}

\newcommand{\yr}     {\hbox{$\,\mathrm{yr}$}}
\newcommand{\cm}     {\hbox{$\,\mathrm{cm}$}}
\newcommand{\gm}     {\hbox{$\,\mathrm{gm}$}}

\onecolumn

\begin{document}

\title[Orbital evolution and growth of protoplanets]
{On the orbital evolution and growth of protoplanets
embedded in a gaseous disc}
\author[J.C.B. Papaloizou and J.D. Larwood]
{J.C.B. Papaloizou and J.D. Larwood\thanks{j.d.larwood@qmw.ac.uk} \\
Astronomy Unit, Queen Mary \& Westfield College, Mile End Road, London E1 4NS}

\date{Received 1999 November 23.}
\volume{000}
\pagerange{\pageref{firstpage}--\pageref{lastpage}}
\pubyear{0000}

\maketitle
\label{firstpage}

\begin{abstract}
We present a new computation of the linear tidal interaction
of a protoplanetary core with a thin gaseous disc in which it is
fully embedded. For the first time a discussion of
the orbital evolution of cores on eccentric orbits
with eccentricity ($e$) significantly larger than the gas-disc scale
height to radius ratio ($H/r$) is given. We find that the
direction of orbital migration reverses for $e>1.1H/r$.
This occurs as a result of the orbital crossing of
resonances in the disc that do not overlap the orbit when the
eccentricity is very small. In that case resonances always
give a net torque corresponding to inward migration.
Simple expressions giving approximate
fits to the eccentricity
damping rate and the orbital migration rate are presented.
We go on to calculate the rate of increase of the mean eccentricity
for a system of protoplanetary cores due to dynamical
relaxation. By equating the eccentricity damping time-scale with the dynamical
relaxation time-scale we deduce that, for parameters thought to be applicable to
protoplanetary discs, an equilibrium between eccentricity damping and
excitation through scattering is attained on a
$10^3${--}$10^4$\yr~time-scale, at $1$\au.
This equilibrium is maintained during the further
migrational and collisional evolution of the system, which occurs on
much longer time-scales. The equilibrium thickness of the
protoplanet distribution is related to the equilibrium eccentricity
and is such that it is generally well confined within the gas disc.
By use of a three dimensional direct summation N-body code we simulate the
evolution of a system of protoplanetary cores, initialised
with a uniform isolation mass of $0.1M_{\oplus}$, incorporating our
eccentricity damping and migration rates.
Assuming that collisions lead to agglomeration, we find that the
vertical confinement of the protoplanet distribution permits
cores to build up in mass by a factor of $\sim10$
in only $\sim10^4$\yr, within $1$\au. The time-scale
required to achieve this is comparable to the migration time-scale.
In the context of our model and its particular initial conditions
we deduce that it is not possible to build up a massive enough core
to form a gas giant planet, before orbital migration 
ultimately results in the preferential
delivery of all such bodies to the neighbourhood of the central star.
It remains to be investigated whether different disc models
or initial planetesimal distributions might be more
favourable for slowing or halting the migration,
leading to possible giant planet formation at intermediate radii.
\end{abstract}


\section{Introduction}

The last decade has seen the discovery of Jupiter-sized extrasolar planets
orbiting Solar-type stars. In roughly one fifth of the cases recorded to date
a planet is inferred to be in nearly circular orbit
(having eccentricities less than $0.05$) at a
distance from the parent star of less than fifteen
Solar radii ($\sim 0.06$\au).
In the rest of the detections the planets are inferred to have
eccentricities up to $0.7$ (with a mean value of $0.3$), for
semimajor axes in the range $0.07${--}$3$\au~(Vogt et al. 2000).
Thus the orbital characteristics of many of the known extrasolar systems
contrast strongly with the gas giants of the Solar system that have
eccentricities less than about $0.05$ and semimajor axes greater than
$5$\au. These new discoveries
pose a considerable challenge to the standard models of planetary
formation that are based on our knowledge of the Solar system alone.

According to the accumulation model, protoplanetary cores, the
building blocks of planet formation, condense from the interstellar
dust grains that are initially suspended in the gaseous protostellar
disc. These first aggregate to form planetesimals of mass
$10^{18}$--$10^{20}$\gm~(e.g. Weidenschilling \& Cuzzi 1993).
Seed cores then grow up to an isolation mass, passing through a process
of runnaway accretion (Lissauer \& Stewart 1993).
The duration of this growth phase is expected to be 
on the order of $10^5$\yr~at $1$\au~(eg. see the reviews
by Lin, Bryden, \& Ida, 1999, and Papaloizou, Terquem \& Nelson, 1999
and references therein). After isolation
the orbits of cores do not overlap so that
the system evolution slows
significantly. Adopting a heavy element surface density of $6\gm\cm^{-2}$, 
characteristic of the minimum mass Solar nebula, the protoplanetary
cores at  isolation are thought to have roughly $0.03M_{\oplus}$
at $1$\au~from the Sun (Lissauer \& Stewart 1993,
Lin et al. 1999). For a surface density ten
times larger this increases to $1M_{\oplus}$.

The cores subsequently interact under their mutual gravitational
attraction which can lead to collisions and the consequent production
of more massive cores. If a critical mass scale $\sim 15M_{\oplus}$
can be reached (e.g. Mizuno 1980, Bodenhiemer \& Pollack 1986), 
gas accretion can be initiated. Without further growth,
the core can build up a
substantial gaseous envelope which can contribute much of the final
mass of the object as in the case of Jupiter and
presumably the giant planets in
extrasolar systems. The time-scale for the
gas accretion process is estimated to be
$10^6$--$10^7$\yr~(e.g. Bodenheimer \& Pollack 1986). A potential problem is
that the time required to build up a core with the critical mass may
become very long if the cores become isolated and do not
migrate (Lissauer \& Stewart 1993).

However, in order to discuss the evolution of protoplanetary
cores, their gravitational interaction with the gas disc must be
taken into account. This interaction produces wave excitation
and angular momentum exchange (Lin \& Papaloizou 1979)
which can lead to orbital migration (Goldreich \& Tremaine 1980).
Low mass protoplanets interact linearly with the disc
and undergo type I inward migration (Ward 1997b).
Massive protoplanets interact nonlinearly and undergo
type II migration (Lin \& Papaloizou 1986). In this paper
we shall be concerned with type I migration.
In addition to orbital migration, eccentricity damping
is produced (Goldreich \& Tremaine 1980, Artymowicz 1993, 1994).
Normally, migration calculations are undertaken assuming the
eccentricity is smaller than the ratio of disc thickness to radius.
Since the Solar nebula is generally believed to have been very thin,
the latter is a small number, making the small eccentricity assumption
very restrictive. Gravitational scattering amongst the cores acts to
pump up their eccentricity until damping effects limit it.
The possibility that the eccentricity is larger than
the ratio of disc thickness to radius should be taken into account.
We present here a new computation of the protoplanet-disc
interaction in which it is calculated by summing over all
resonances required to ensure validity, for eccentricities up to five times
the ratio of disc thickness to radius, although in principle any value 
smaller than unity could be considered.

We use our results to obtain an estimate for the equilibrium
values of the mean eccentricity and vertical thickness for a
system of protoplanets.
This is obtained by balancing pumping through gravitational
scattering with damping through tidal interaction with
the nebula. The effects of gravitational scattering and
the details of the equilibrium
are obtained by use of the Boltzmann ${\cal H}$ theorem
applied to the Fokker-Planck equation.
The limitation of eccentricity because of the
protoplanet-disc interaction is manifest in the restricted
vertical extent of the protoplanet swarm. That is found 
to be always well confined within the gaseous disc,
which in turn stabilises the collision rate and promotes rapid core
agglomeration in comparison to the gas-free case. We investigate
the consequences of this through numerical simulations.

In Section $2$ we calculate the dynamical relaxation time-scale for a system
of equal mass protoplanetary cores by
use of the Boltzmann ${\cal H}$ theorem applied
to the Fokker Planck equation. In Section $3$ we derive
the eccentricity damping and orbital migration time-scales
in the linear regime for
a protoplanetary core that is fully embedded in a thin gaseous disc
with surface density $\propto r^{-1.5}$.
In Section $4$ we apply these results to find
the vertical extent of a protoplanet swarm at equilibrium
as a function of the nebula properties. In Section $5$ we present
numerical tests of our analysis using a direct summation N-body code
in combination with an implementation of our nebula torque model.
Specifically, we consider an ensemble of one hundred $0.1M_{\oplus}$ cores
distributed interior to $1$\au.

The general finding is that, for characteristic
protoplanetary disc parameters, because the eccentricity damping
time-scale is significantly shorter than the migration
time-scale, a quasi-equilibrium distribution is obtained in a 
$10^3${--}$10^4$\yr~time-scale at $1$\au~and
which is otherwise proportional to the local disc radius.
 Inward migration occurs on the
much longer migration time-scale and aids the accumulation of
up to earth-mass cores in the inner regions of the disc
on a $10^4${--}$10^5$\yr~time-scale at $\sim 1$\au. However,
these objects undergo rapid
orbital migration towards the central star on the same time-scale, thus
in this particular model, gas accretion onto a sufficiently
massive core to make a giant planet 
within the disc lifetime requires a mechanism to halt the migration.
One possibility is termination of the disc due to a magnetospheric
cavity (Lin, Bodenhiemer \& Richardson 1996).
On the other hand if cores of the same initial
mass are formed at much larger radii, they may survive for the disc lifetime
without closely approaching the central star.
In Section $6$ we summarise and discuss these findings.
For a first reading of the manuscript we suggest that the general
reader moves directly to Section $4$.

\section{Analysis of Gravitational Scattering}
\label{sec:VGR}

A system of many bodies interacting under their mutual gravitation
can be described by the Fokker-Planck equation which we write in the form:

\beq
{{\rm D}f_{\alpha}\over {\rm D}t}=
\Gamma_{coll}(f_{\alpha}) +\Gamma_{gas}(f_{\alpha})
\label{FPEQ}.
\eeq

\noindent Here $f_{\alpha}$ denotes the phase space number density of bodies
with mass $m_{\alpha}$. The operator giving evolution due to
gravitational scattering is $\Gamma_{coll}$, and that giving evolution due to
interaction with the gaseous disc is $\Gamma_{gas}$.
The latter combines the effects of orbital migration,  and eccentricity
and inclination  damping.
The derivative operator is taken following a particle orbit considering
gravitational forces due to the central mass such that

\beq
{{\rm D}\over {\rm D}t} \equiv {\partial \over \partial t} + 
{\bf v}\cdot {\partial \over \partial {\bf r}}
-{\nabla \Phi}\cdot{\partial \over \partial {\bf v} } .
\eeq

\noindent Where we refer to the central potential $\Phi = -GM_{*}/r$,
with the particle position and velocity vectors being denoted by ${\bf r}$ and
${\bf v}$ respectively. The central mass is $M_{*}$ and $r = |{\bf r}|$.

\subsection{Local shearing sheet approximation}

As a result of the large relative velocities occuring between objects
widely separated in radius,
effective gravitational scatterings will occur on a radial length scale
comparable to the vertical thickness of the planetesimal disc.
The smallness of the latter quantity suggests use of the
Goldreich \& Lynden-Bell (1964) shearing sheet approximation.
In this approximation we consider a uniformly rotating local
Cartesian coordinate system in which the origin, located
at some point of interest in circular orbit with
radius/semimajor axis $r_0$, corotates
with the Keplerian angular velocity $\Omega$. The $x$-axis
points radially outwards and the $y$-axis points in the
azimuthal direction while the $z$-axis points in the
vertical direction. A linear expansion for the central potential
is used such that

\beq
\nabla \Phi = (\Omega^2 r_0 -2\Omega^2 x, 0, \Omega^2 z).
\eeq

In this scheme we have for an axisymmetric disc with accordingly
no dependence on $y$,

\beq
{{\rm D} f_{\alpha} \over {\rm D}t} \equiv
{\partial f_{\alpha} \over \partial t}
+ v_x {\partial  f_{\alpha}\over \partial  x}
+ v_z {\partial f_{\alpha} \over \partial  z}
- 2\Omega v_x{ \partial f_{\alpha}  \over \partial v_y }
+ (3\Omega x+2v_y)\Omega { \partial f_{\alpha} \over \partial v_x }
-\Omega^2 z{ \partial f_{\alpha}\over \partial  v_z } ,
\label{CBE}
\eeq

\noindent where ${\bf v} = (v_x,v_y,v_z)$ and ${\bf r} = (x,y,z)$.
The Liouville equation ${\rm D} f_{\alpha} / {\rm D}t =0$ has from
Jeans' theorem general solutions
with $f_{\alpha}$ being an arbitrary function of the integrals of the motion.
Later we shall adopt such a solution, corresponding to an anisotropic Gaussian:

\beq
f_{\alpha} = C_{\alpha}
\exp\left( -{v_x^2\over 2\sigma_x^2} -{u_y^2\over 2\sigma_y^2}
-\frac{v_z^2 +\Omega^2 z^2}{2 \sigma_z^2}\right)
\label{VelI}.
\eeq

\noindent Here the velocity dispersions
$(\sigma_x,\sigma_y,\sigma_z)$ are constant
but such that $\sigma_y =\sigma_x /2$. There is no constraint on $\sigma_z$.
The velocity $u_y =v_y +3\Omega x/2$ is measured relative to the
local circular velocity, $v_y = -3\Omega x/2$, and we shall adopt
local relative velocity vectors ${\bf v} = (v_x, u_y, v_z)$.
The constant $C_{\alpha}$ is related to the spatial number density
in the midplane, $n_{\alpha},$
through

\beq
C_{\alpha}  = { n_{\alpha}  \over (2 \pi) ^{3/2}
\sigma_x   \sigma_y  \sigma_z }.
\eeq 

\subsection{System evolution through gravitational scattering}

Encounters between planetesimals that occur without direct physical impacts
tend to convert kinetic energy from shear into that of random motions,
much as viscosity converts energy from shear into thermal motions
in a gaseous disc. In the situation considered here
evolution due to scattering occurs on a time-scale much longer than
orbital. To investigate this phenomenon we use the form of
$\Gamma_{coll}$ given by Binney and Tremaine (1987).
This applies to a homogeneous system and so neglects rotation
about the central mass. This should be reasonable so long as the time-scale
associated with an encounter is short compared to $\Omega^{-1}$.
This in turn requires that $\sigma_x/\Omega > r_H$, where $r_H$
is the Hill radius $r_0(\frac{m_{\alpha}}{3M_{*}})^{1/3}$,
appropriate to the characteristic mass $m_{\alpha}$.

Following Binney and Tremaine (1987), we write using the summation convention
for repeated indices

\beq
\Gamma_{coll}(f_{\alpha}) = 
-\frac{\partial}{\partial v_i}(A_i f_{\alpha})
+\frac{1}{2}\frac{\partial}{\partial v_i}
 \left( D_{ij} \frac{\partial f_{\alpha}}{\partial v_j} \right)
\label {COP}.
\eeq

\noindent Here

\beq
A_i = 4\pi G^2 \ln(\Lambda) m_{\alpha}^2{\partial h
 \over \partial v_i}
\eeq

\noindent and

\beq
D_{ij}=  4\pi G^2 \ln(\Lambda) m_{\alpha}^2{\partial^2 g
 \over \partial v_i \partial v_j},
\eeq 

\noindent with

\beq
(g,h)= \left( 
\int f_{\alpha}({\bf v'})|{\bf v}-{\bf v'}|
\,{\rm d}^3{\bf v}',
\int {f_{\alpha}({\bf v'})\over |{\bf v}-{\bf v'}|}
\,{\rm d}^3{\bf v}'\right). 
\eeq

\noindent For $\Lambda$ we take
$\Lambda = 3\sigma_x^2 H_\alpha /(4Gm_{\alpha})$,
giving the ratio of maximum to minimum impact parameters as disc semi-thickness
to impact parameter for a typical deflection (e.g. Binney \& Tremaine 1987).
The above formalism applies to one species of planetesimal with mass
$m_{\alpha}$. Generalization to include a system of interacting
planetesimals with different masses is straightforward. However,
for simplicity we shall assume all the planetesimals
have equal mass in the analysis presented here.

\subsection{Growth of the velocity dispersion}

The effect of gravitational scattering is to cause the velocity
dispersion to increase. This is most easily seen by formulating
the Boltzmann ${\cal H}$ theorem. This states that for a single mass species

\beq
{\cal H}\equiv -\int f_{\alpha}\ln(f_{\alpha})\,{\rm d}^3{\bf v}
\eeq

\noindent increases monotonically with time.
${\cal H}$, which can be related to the entropy,
can remain constant only for an isotropic Gaussian which cannot be
attained here because of the form of particle orbits in the central
potential (see equation \ref{VelI}). Using the distribution function given
by (\ref{VelI}) we obtain at the midplane ($z=0$),
 assuming that $n_{\alpha}$ is constant
and for fixed ratio of velocity dispersion components,

\beq
{{\rm d}{\cal H}\over{\rm d}t} =
{ 3 n_{\alpha} \over \sigma_x} {{\rm d}\sigma_x  \over {\rm d}t}.
\label{Hdot}
\eeq

\noindent Thus the velocity dispersion increases with time when
${\rm d}{\cal H}/{\rm d}t > 0.$
Here we make the assumption $n_{\alpha}$ is constant and specialize
to the midplane where $z=0.$ We comment
that almost identical results are obtained if instead a vertical
integration is done and the surface density is assumed constant.

Evaluating ${\rm d}{\cal H}/{\rm d}t$, neglecting for the time
being the effects of migration and damping, we obtain using (\ref{COP}):

\beq
{{\rm d}{\cal H}\over {\rm d}t} =
-4\pi G^2 m_{\alpha}^2\ln(\Lambda)
\int { {\partial f_{\alpha}({\bf v})
 \over \partial v_i}  {\partial f_{\alpha} ({\bf v'})
 \over \partial v_i^\prime}\over |{\bf v} - {\bf v}'|}
\,{\rm d}^3{\bf v} \,{\rm d}^3{\bf v'}
+ 2\pi G^2 m_{\alpha}^2\ln(\Lambda)
\int \left( { {\partial f_{\alpha} ({\bf v})
 \over \partial v_i}  {\partial f_{\alpha} ({\bf v})
 \over \partial v_j} f_{\alpha} ({\bf v'})\over f_{\alpha} ({\bf v})}\right)
 \left({\partial^2 |{\bf v} - {\bf v}'|
 \over \partial v_i \partial v_j}\right)
\,{\rm d}^3{\bf v} \,{\rm d}^3{\bf v'}
\label {CAP}.
\eeq

Note that the two terms on the right hand side of (\ref{CAP}) give
contributions of opposite sign. However, a net positive result
is guaranteed by the ${\cal H}$ theorem which is why we follow that approach.
The integrals in (\ref{CAP}) are most easily performed by
use of Fourier transforms and the convolution theorem with the result that 
at $z=0.$
\beq
{{\rm d}{\cal H}\over {\rm d}t} = 
(4\pi)^2 G^2 m_{\alpha}^2\ln(\Lambda)(C_{\alpha}
\sigma_x \sigma_y \sigma_z)^2  
\int \exp(-k_i^2\sigma_i^2)
\left( {k_i^2\over k^4\sigma_i^2} -2\right)\,{\rm d}^3{\bf k}.
\eeq  

\noindent When $\sigma_y=\sigma_z$, the integral may be evaluated analytically.
Using (\ref{Hdot}) we then obtain for the relaxation time,
$t_R = \sigma_x/({\rm d} \sigma_x/ {\rm d}t)$,

\beq
{1\over t_R}=
 {8(\pi)^{1/2} G^2 m_{\alpha}^2\ln(\Lambda) n_{\alpha}\over \sigma_x^3}
\left[{\sqrt{3}\over 4}\ln\left({2+\sqrt{3} \over
2-\sqrt{3}}\right)-1\right].
\label{tr}
\eeq

\noindent Here we have also specialized to the Keplerian case for which
$\sigma_y =\sigma_x /2$. 

In general
the vertical velocity dispersion would be affected by inclination damping.
In the case of no damping we expect collisional effects to cause
evolution towards isotropy such that
$\sigma_z^2 =(\sigma_x^2 + \sigma_y^2)/2$, or $\sigma_z/\sigma_x = 0.79$.
Damping reduces this value, but as long as the rate of inclination
damping is comparable to or  less than that for eccentricity damping (as is
expected to be the case from the results of Ward \& Hahn 1994), this effect
is not very significant (see below).
To obtain a  consistent estimate from (\ref{tr}) we adopt the value of
$\sigma_z/\sigma_x = 0.5$. We shall express the relaxation time
in terms of the semi-thickness of the planetesimal disc
$H_{\alpha}=\sigma_z/\Omega$ and the planetesimal
surface density $\Sigma_{\alpha} =\sqrt{2\pi}H_{\alpha} m_{\alpha}n_{\alpha}$.
Under these approximations we now have:

\beq
\Lambda =
\left( \frac{3M_*}{m_\alpha} \right)
\left( \frac{H_\alpha}{r_0}   \right)^3 .
\label{lambda}
\eeq

\noindent Thus our analysis is expected to break down when the disc thickness
is smaller than the protoplanet Hill radius. Here we obtain:

\beq
{1\over t_R} = 0.03\ln(\Lambda){M_D m_{\alpha}\over M_*^2}
\left({r_0\over H_{\alpha}}\right)^4\Omega,
\label{trr}
\eeq

\noindent where $M_D =\pi \Sigma_{\alpha}r_0^2$ gives an estimate of
the protoplanet-disc mass within $r_0$. Adopting values consistent
with our numerical work: $M_*=1M\sun$,
$m_{\alpha} =0.1M_{\oplus}$, $M_D=10M_{\oplus}$, and
$H_{\alpha}/r_0 =0.01$, we find for $r_0 =1$\au~that
$t_R\sim 3000$\yr. We comment that with $\sigma_x/\sigma_z=2$,
$H_{\alpha}/r_0 =0.01$ corresponds to a mean eccentricity of
$2\sqrt{2} H_\alpha/r_0 \sim 0.03$.

From this discussion we expect an equilibrium velocity
distribution characterized as above,
to have been obtained in a time $\sim 10^3$\yr~within $1$\au,
which is consistent with the numerical results presented below.
The actual equilibrium levels of the velocity dispersion
are obtained by balancing pumping due to gravitational scattering
against damping due to tidal interaction with the nebula.
We calculate the latter in the next section.

\section{Tidal torques resulting from nebula interaction}
\label{TORQ}
Here we calculate the tidal interaction of a protoplanetary core
with the gaseous disc. We assume the protoplanet mass is small
enough that the disc response is linear. Thus we deal with migration
of type I (Ward 1997b).

We employ a simple two dimensional model disc for which the gas surface 
density $\sigma \propto r^{-3/2}$, and
the sound speed $c \propto r^{-1/2}$. The putative aspect ratio
$H/r$ is thus independent of radius and corotation torques
are absent (e.g. Korycansky and Pollack 1993).
The gas angular velocity $\Omega$ is then given by

\beq
\Omega^2 = {GM_*\over r^3}\left(1 - {5 c^2 r\over 2GM_*}\right),
\eeq

\noindent and as for Kepler's law this is proportional to $r^{-3}$.

The protoplanet or planetesimal of mass $m_{\alpha}$ exerts
a tidal potential 

\beq
\Psi = -{Gm_{\alpha}
\over \sqrt{r^2 + R^2 -2rR\cos(\varphi-\varphi_{\alpha})+b^2}},
\eeq

\noindent where $(R, \varphi_{\alpha})$ are the cylindrical coordinates
of the protoplanet on its general Keplerian orbit.
To prevent divergences we incorporate a softening parameter $b$.
This can be thought of as taking into account the disc vertical
thickness.

The protoplanet potential is written as a double Fourier series
(e.g. Goldreich and Tremaine 1978) in the form

\beq
\Psi =\Re \left(\sum_{m=0}^{\infty}\sum^{\infty}_{n= -\infty}
\Psi_{n,m}\exp{i[(n-m)\omega t +m\varphi]}\right).
\eeq

\noindent Here $\Re$ denotes that the real part is to be taken
and $\omega$ is the orbital frequency of the protoplanet
taken to have orbital eccentricity $e$ and semimajor axis $a \equiv r_0$.
The Fourier coefficients are given by

\beq
\Psi_{nm} = {\omega\over 2\pi^2(1+\delta_{m,0})}
\int^{2\pi/\omega}_0 \int^{2\pi}_0  
\Psi \cos[m(\varphi -\varphi_{\alpha})]d\varphi
\exp{[-i(n-m)\omega t-im\varphi_{\alpha}]}\,{\rm d}t .
\eeq

\noindent Where $\delta$ is the Kronecker delta.
Note that we do not include the indirect term in the above
as the contribution  due to this is much smaller than that
due to terms with large $m$. When $e=0$, $\Psi_{nm}=0$
for all $n \ne 0$.

In the linear regime the torques due to each
Fourier component may be evaluated separately and the results 
summed. The Fourier component $(n,m)$ produces a tidal disturbance
rotating with a pattern speed $\Omega_p =(m-n)\omega /m$.
In our model disc only Lindblad resonances are important
and if they exist these occur at  the two locations in the disc where

\beq
m^2(\Omega_p -\Omega)^2 =
\left[(m-n)\omega - m\Omega \right]^2 =
\kappa^2 + {c^2m^2\over r^2}.
\eeq

\noindent For $m > 0$,
the outer Lindblad resonance with $\Omega < \Omega_p$
occurs where

\beq
m(\Omega_p -\Omega) =
\left[( m-n)\omega -m\Omega \right] =
\sqrt{\kappa^2 + {c^2m^2\over r^2}}.
\eeq

\noindent The inner Lindblad resonance with $\Omega > \Omega_p$
occurs where

\beq
m(\Omega_p -\Omega) =
\left[( m-n)\omega -m\Omega \right] =
-\sqrt{\kappa^2 + {c^2m^2\over r^2}}.
\eeq

Here the epicyclic frequency $\kappa =\Omega$.
Note the term $c^2m^2/r^2$, normally neglected 
for small $m$, leads to an accumulation of resonances
at the two locations
where $(\omega-\Omega) = \pm c/r$ as $m \rightarrow \infty$
for finite $n$. These locations are at a distance $2H/3$
inside and outside the semimajor axis of the 
orbit. There are no closer resonances 
for large $m$. This leads to the well known torque cut off
for $m\gg r/H$ (Goldreich \& Tremaine 1980, Artymowicz 1993) which
results in convergence of torque sums in the case of a circular
protoplanet orbit $(n=0)$, even when gravitational softening is omitted.
However, for general $n$, resonances 
may approach the coorbital radius and so results become more sensitive
to the softening parameter. For our disc model the
resonant locations occur where

\beq
\Omega={( m-n)\omega\over
m \pm \sqrt{ 1+ (H^2m^2/r^2)}}.
\eeq

\noindent Thus a strictly coorbital resonance occurs if
$n = \pm \sqrt{ 1+ (H^2m^2/r^2)}$. Even when
$|n|=1$ two resonances occur
at locations satisfying $(\Omega -\omega)\sim (H/r)\omega$
when $m \sim r/H$. These have been found to lead to rapid
eccentricity damping (e.g. Artymowicz 1994).
Up to now terms with $|n| > 1$ have not been considered, which
is reasonable for very small $e$.
Here we consider all $n$ necessary for convergence. Once $e > H/r$,
significantly larger values of $n$ are required and these may produce
more closely coorbital resonant effects.

We comment that for $n=0$, the inner Lindblad resonances
are all located interior to the semimajor axis while the outer
Lindblad resonances are all located exterior to it.
This results in the inner disc unambiguously providing
a positive torque and the outer disc unambiguously providing
a negative torque. However, when $n > 0$, inner/outer Lindblad resonances
may fall external/internal to the orbital semimajor axis. This means
that the interior or exterior discs may produce torques
of either sign on an orbit with significant eccentricity.
Physically, a protoplanet at apocentre may rotate
more slowly than coorbital disc material. If $e$ is large enough
compared to $H/r$ the exterior disc can then  exert a positive torque.
We apparently find that this phenomenon may lead to a net
torque reversal for sufficiently eccentric orbits. 

\subsection{Rate of orbital evolution}

To evaluate the rate of change of protoplanet angular momentum
arising from the $(n,m)$ Fourier component,
we use the Goldreich and Tremaine (1978) torque formula
as modified by Artymowicz (1993) (see also Ward 1997b) in the form

\beq
{{\rm d}J_{n,m}\over {\rm d}t} = {Sr^2\Sigma\over 3\Omega \Omega_p}
\left[
r{{\rm d}\Psi_{n,m}\over {\rm d}r} +
{2 m^2(\Omega -\Omega_p)\Psi_{n,m}\over \Omega }
\right]^2
\left[1\over1+4m^2c^2/(r^2\Omega^2)\right]
\label{GT}.
\eeq

Here $S=1$ for an inner Lindblad resonance that transfers angular momentum to
the planet, $S=-1$ for an outer Lindblad resonance
that removes it from the planet and the right hand side of (\ref{GT})
is to be evaluated at the resonant location.
The associated rate of change of orbital energy is then
related to the rate of change of angular momentum through:

\beq
{{\rm d} E_{n,m}\over {\rm d}t} = 
\Omega_{p,n,m} {{\rm d} J_{n,m}\over {\rm d}t},
\eeq

\noindent where we have added the subscripts $(n,m)$ to the
pattern speed $\Omega_p$ to denote that it is associated with
the corresponding Fourier component.

By summing over $(n,m)$ we find the total rate of change of 
angular momentum ${\rm d}J/{\rm d}t$ and we define the orbital
migration time $t_{m}$ such that

\beq
t_{m} = -\frac{J}{\left({\rm d}J/{\rm d}t\right)},
\eeq

\noindent where $J$ is the angular momentum of the protoplanet.
Migration here is thus defined in terms of the total torque
exerted on the orbit. The migration time so defined is positive
when the total torque is negative.

Similarly we find the total rate of change of orbital energy
and then the rate of change of eccentricity using

\beq
{1\over J}\left({{\rm d}E\over {\rm d}t}-\omega {{\rm d}J\over {\rm d}t}\right)
= \omega {1\over J}{{\rm d}J\over {\rm d}t}
\left[{(1-\sqrt{1-e^2} )\over \sqrt{1-e^2}}\right]
+\omega{e\over (1-e^2)^{3/2}}{{\rm d}e\over {\rm d}t}.
\eeq

\noindent The eccentricity evolution time-scale is then

\beq
t_e = \frac{e}{|{\rm d}e/{\rm d}t|}.
\eeq

\noindent When, as is always found to be the case here,
${\rm d}e/{\rm d}t<0$, $t_e$ gives the circularisation time-scale.

\begin{figure*}
\centerline{\epsfig{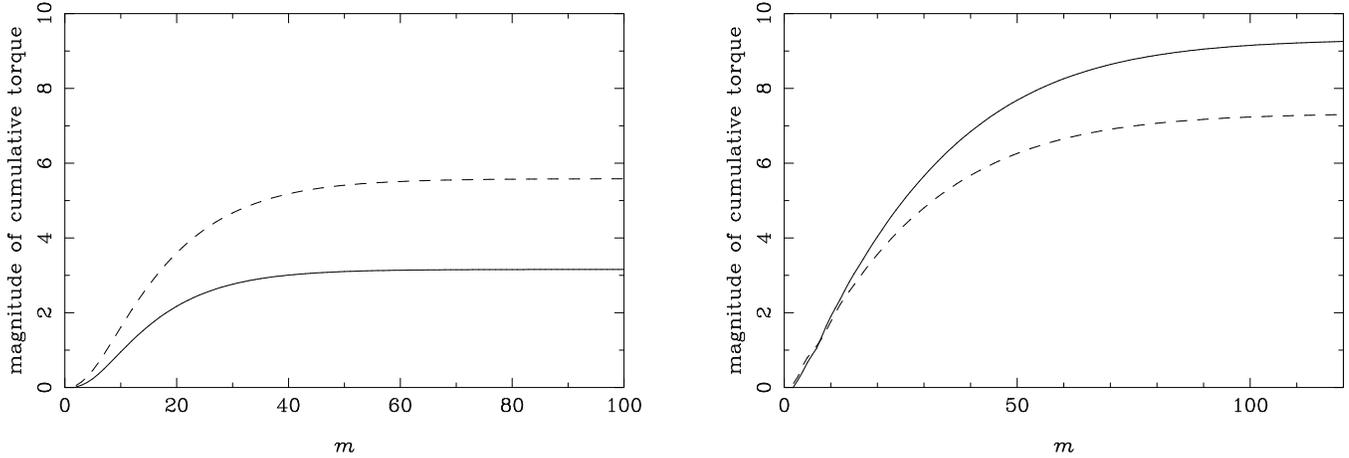}}
\caption{The magnitude of the cumulative torque in arbitrary units as
a function of $m$. The net torque contributions arising from inner
Lindblad resonances are plotted with solid
lines and those arising from outer Linblad resonances are plotted
with dashed lines.
In the left-hand panel we show the magnitudes of the net torque
contributions for very small $e$ and $H/r=0.07$.
In this case only $n=0$ contributes for each $m$.
The outer Linblad resonances predominate leading to net inward migration.
In the right-hand panel we show the magnitudes of the net torque
contributions for $e=2H/r$ and $H/r=0.07$. In this case the torque at each 
$m$ has all of the significant contributions from each value of $n$
included. In this case, unlike when $e$ is small,
the inner Lindblad resonances predominate, leading to net outward
migration. Note that these resonances are not all located in the inner disc.}
\label{fig1}
\end{figure*} 

We have performed calculations of $t_{m}$
and $t_e$ taking into account as many resonances as required for convergence.
We comment that we did not make any large $m$ asymptotic approximation
as this was found to affect the results significantly.
We illustrate the contributions to the torque for very small $e$
for which only $n=0$ and $n=1$ need to be considered in the left-hand
panel of Figure $1$.
The calculation illustrated is for $H/r =0.07$.
In this case the contribution from the outer Lindblad
resonances that are located in the outer disc dominates and the
migration is inwards. The eccentricity is always damped.

We found that departures from the small $e$ case occured once
$e > H/r$.
The sign of the torque actually reverses once $e$ exceeds $\sim 1.1H/r$.
Significant changes occur
because for such values the radial excursion 
causes the protoplanet to cross 
the circular orbit resonances a distance $2H/3$
away from the orbital semimajor axis.

In the right-hand panel of Figure $1$ we show the torque
contributions from the inner and outer
Lindblad resonances for $H/r=0.07$ and $e=2H/r$.
In this case the inner Lindblad resonances
dominate giving a net positive torque.
But these are not always located at radii interior to the
orbital semimajor axis. The eccentricity is always damped,
but note that $t_e \propto e^3$ for $e > H/r$, so that the more eccentric
orbits feel very much weaker damping than the near circular ones.

We found that the dependence on the gas-disc thickness was such that
$t_{m} \propto (H/r_0)^{2}$, and $t_e \propto (H/r_0)^{4}$,
with $H$ now being evaluated at $r=r_0$.
This scaling, together with the coorbital resonance contributions
to $t_e$, result in $t_e$ being much shorter than $t_{m}$.
Accordingly we expect an equilibrium eccentricity distribution
to be set up in a system of gravitationally interacting
protoplanets embedded in a gas disc, assuming that they are
massive enough so that tidal torques dominate
over those due to gas drag. After this equilibrium is set up,
longer time-scale evolution is expected to occur
due to orbital migration and physical collisions.

We also note that the torque results were dependent on the softening
parameter used with $t_{m}$ and $t_e$, scaling as $b^{1.75}$ and
$b^{2.5}$ respectively, for $b$
in the range $0.4H${--}$H$. This leads to an uncertainty reflecting
the inadequacy of using a two dimensional flat disc model as the
softening would occur naturally in a fully
three dimensional treatment.
Fortunately our simulations of protoplanet discs
are insensitive to this issue. Also as we find the eccentricities
are for the most part limited to small values, making
the results for this case 
the most important  for the work presented here.
For small $e$ our results are broadly consistent with those of Ward (1997b)
and Artymowicz (1993, 1994).
From our calculations we found the approximate fits to
$t_{m}$ and $t_e$ given by

\beq
t_{m} = 3.5\times 10^5 f_s^{1.75}
\left[\frac{1+\left({er_0\over 1.3H}\right)^5}
{1-\left({er_0\over 1.1H}\right)^4}\right]
\left({H/r_0\over 0.07}\right)^2
\left({ 2M_{J}\over M_{GD}}\right)
\left({ M_{\oplus}\over m_{\alpha} }\right)
\left({r_0\over 1\au}\right)\yr
\label{mfit}
\eeq

\noindent and

\beq
t_e = 2.5\times 10^3 f_{s}^{2.5}
\left[1+\frac{1}{4}
\left({e \over H/r_0 }\right)^3\right]
\left({H/r_0\over 0.07}\right)^4
\left({ 2M_{J}\over M_{GD}}\right)
\left({ M_{\oplus}\over m_{\alpha} }\right)
\left({r_0\over 1\au}\right)\yr.
\label{efit}
\eeq                              

\noindent Here the gas disc
has a mass $M_{GD}$ contained within $5$\au.  At other radii
the mass of the gas disc 
is assumed to scale as $r_0^{1/2}$, as implied by
the model surface density. 
The Jovian mass is $M_J$.
We note that the factor due to softening,
$f_s\equiv\left({2.5b\over H }\right)$,
is hereinafter taken as unity throughout, coressponding to
$b=0.4H$.
For $H=0.07r_0$, $M_{GD}=20M_J$,
and $m_{\alpha} = 0.1M_{\oplus}$, we obtain
$t_e \sim 10^3$\yr~and $t_m \sim 10^5$\yr,
in the limit of $e\ll H/r$ at $1$\au.

\subsection{The contribution from inclination damping}

In estimating the damping forces we return to the Boltzmann
${\cal H}$ theorem and consider the contribution
to ${\rm d}{\cal {\cal H}}/{\rm d}t$ one obtains from
$\Gamma_{gas}(f_{\alpha})$
in equation (\ref{FPEQ}). We suppose this can be represented
through the action of a body force ${\bf F}$ per unit mass
such that

\beq
\Gamma_{gas}(f_{\alpha})=-{\partial \over \partial {\bf v}}
\left( {\bf F} f_{\alpha}\right).
\eeq

\noindent Here if ${\bf F} =-( 2v_x/t_e, 0, 2v_z/t_i)$,
the eccentricity and inclination damping times will be
$t_e$ and $t_i$, respectively. As migration occurs on a longer time-scale,
we shall neglect it when considering the velocity dispersion equilibrium.
The damping force ${\bf F}$ gives a contribution to the rate
of increase of ${\cal H}$ in the midplane:

\beq
{{\rm d}{\cal H}\over {\rm d}t}=
-2n_{\alpha}\left({1\over t_e}+ {1\over t_i}\right).
\eeq

\noindent Thus, as long as $t_i\ga t_e$, the equilibrium condition
${\rm d}{\cal H}/{\rm d}t =0$ can be approximated by equating $t_e$
to the relaxation time $t_R$, see equation (\ref{trr}).

\section{Velocity dispersion equilibrium}

As we have seen in Section \ref{sec:VGR}, the effect of
gravitational scattering not involving direct collisions is
to increase velocity dispersion on a time-scale $t_R$. An equilibrium
is attained by balancing this rate of increase against damping due to tidal
interaction of the protoplanet with the gaseous disc. As described in
the previous Section we can approximate damping effects by
ignoring inclination damping, occurring on the time-scale $t_i$,
and only consider eccentricity damping,
occurring on the time-scale $t_e$, provided $t_e\la t_i$.

The time-scale $t_e$ is obtained from the protoplanet-disc
interactions calculated above. As this time-scale is significantly
shorter than the migration time-scale $t_m$, we expect that a
quasi-equilibrium is set up before significant migration
occurs and that this then drives the evolution of the system,
which will occur under conditions of quasi-equilibrium.
To calculate the quasi-equilibrium we use our fit to $t_e$ given by
(\ref{efit}) above and recall our result on the dynamical relaxation time-scale
for a central Solar mass point potential:

\beq
t_R =
\frac{5}{\ln(\Lambda)}
\frac{M\sun^2}{M_D m_\alpha}
\left(\frac{H_{\alpha}}{r_0}\right)^4
\left(\frac{r_0}{1\au}\right)^{3/2}\yr,
\eeq

\noindent where $\Lambda$ is given by equation (\ref{lambda}).
Equating $t_e$ from equation (\ref{efit}) and $t_R$,
assuming $e<H/r_0$ gives the equilibrium semi-thickness of the
protoplanet distribution $H_\alpha$ in terms of the gas disc
semi-thickness $H$:

\beq
\frac{H_{\alpha}}{H} = 0.6 [\ln(\Lambda)]^{1/4}
\left({M_D \over M_{GD}}\right)^{1/4}
\left({r_0\over 1\au}\right)^{-1/8}.
\label{EQUIL}
\eeq

\noindent $H_{\alpha}/H$ depends only very weakly on $\Lambda$ and
therefore (\ref{EQUIL}) is essentially independent of the
protoplanet mass $m_{\alpha}$, and depends only
weakly on other parameters. For
$m_{\alpha} =0.1M_{\oplus}$, protoplanet disc mass
$M_D=10M_{\oplus}$, $H_{\alpha}/r_0 =0.01$, and gas disc mass within
$5$\au~$M_{GD} =20M_J$, we determine $H_{\alpha}= 0.15H$ at $1$\au.
Hence the protoplanet swarm is expected to remain thin
and confined within the gaseous nebula. The characteristic eccentricity
of the equilibrium (see Section $2$) is then roughly
$2\sqrt{2} H_\alpha/r \sim 0.4H/r$.

\section{Numerical Modelling}

In order to investigate the consequences of our approximate analysis,
we have developed a three dimensional direct summation N-body
code to model the gravitational dynamics of a system of protoplanetary
cores. For the time integration we employ a fifth order accurate
Runge-Kutta-Fehlberg scheme with time step control through the estimated
error in velocity magnitude (Press et al. 1992). The scheme is
not strictly conservative with the result being orbital decay in the
circular orbit two body problem. The accuracy tolerance - used in
calculating the computational timestep - was calibrated
through experiments with a one Jupiter mass body in circular orbit about
the Sun at $0.1$\au. We set the accuracy tolerance such that the orbiting
body suffered $0.01$\% orbital decay in $10^4$\yr, being a much weaker
effect than orbital migration due to nebula interaction.

To incorporate the effects of eccentricity and inclination damping and orbital
migration owing to tidal interaction with a gaseous disc we
implemented the following expressions as contributions to the
total acceleration for each particle:

\beq
{\bf a}_{mig} = -\frac{{\bf v}}{t_m} ,
\eeq
\beq
{\bf a}_{damp} = -2\frac{({\bf v}\cdot{\bf r}){\bf r}}{r^2 t_e} 
-2\frac{({\bf v}\cdot{\bf k}){\bf k}}{t_i}.
\eeq

\noindent Here ${\bf k}$ is the unit vector in the vertical direction.
As long as $t_i$ is not significantly
less than $t_e$, tests confirm that inclination damping
does not change the velocity dispersion equilibrium significantly (see below).
Hence the root-mean-square velocity dispersion of the particle distribution
is essentially controlled by the quasi-equilibrium
eccentricity attained by balancing pumping through gravitational
scattering with damping owing to tidal interaction with the gaseous disc.

In all our numerical tests we consider $100$ equal mass particles 
of $0.1M_\oplus$ each, and a centralised Solar mass. The particle
distribution is initialised in the gas-disc midplane such that the
surface density is proportional to $r^{-3/2}$. The positions are
otherwise chosen randomly in the radial range $0.3${--}$1$\au~such that
the separations are greater than twenty Hill radii. The initial
velocity components
in the midplane correspond to Keplerian circular motion about the
central object and vertical velocities are defined as $iv_\phi$, where $i$ is
chosen at random for each particle to lie between $\pm 0.01$, and $v_\phi$
is the Keplerian circular velocity.

In a real protostellar accretion disc we expect the mass of the
disc at a fixed radius to decrease with time on the global viscous
time-scale, which at $1$\au~is typically $\sim 10^4$\yr~(Papaloizou
\& Terquem 1999). We note that
$t_e \propto H^4/M_{GD}$ and $t_m \propto H^2/M_{GD}$,
which tend to increase with diminishing gas-disc mass.
However, this increase is compensated for by the decreasing $H$ found
in nebula evolution models (Papaloizou \& Terquem 1999).
Here we consider a fixed
value $M_{GD}= 20M_J$ which corresponds 
to an early stage in the life of the nebula at an age  
$\sim 10^4${--}$10^5$\yr, in our numerical models.
The gas disc then has a mass $\sim 0.1M_{\odot}$ within a radius $100$\au.
However, our calculations of the equilibrium
eccentricity distribution (see below) can be scaled to a lower
gas-disc mass with smaller
vertical scale height by scaling $H \propto M_{GD}^{1/4}$.
They can also be scaled to apply to larger radii by the scaling transformation:
$(r\rightarrow \lambda r, t\rightarrow \lambda^{1.5}t,
M_{GD}\rightarrow M_{GD}\lambda^{-1/2})$.
Thus all radii can be multiplied by $9$ provided all times
are multiplied by $27$ and the disc mass within $5$\au~is reduced by a
factor of $3$.

In some of our numerical models we allow the masses to collide and agglomerate.
We simulate this process by replacing closely approaching particle pairs
with a single mass occupying the centre of mass position of the original
pair: the new masses are set up to conserve the mass and linear momentum
of the original particle pairs. Binary agglomeration occurs when the separation
is less than the sum of the radii associated with each component.
For this purpose each mass is assigned a radius that gives a mean
density equal to the lunar value. This approach assumes that the
relative motion of the colliding pair is completely dissipated
and that the liberated kinetic energy
does not exceed the total binding energy of the putative bodies.
For the relatively large mass scales considered here the latter
is expected to be realistic (Lecar \& Aarseth 1986). Furthermore, modelling of
collisional fragmentation shows that $>90\%$ of the total mass
remains in the merged core and the existence of remnant ejecta
is not significant in the subsequent evolution of the planetary
system (Alexander \& Agnor 1998), and thus can be ignored.

We note that if agglomeration is included in the simulations as
described above, and the $\lambda$ transformation is applied,
the physical radius of cores is also scaled. This has the affect
that for example if $\lambda >1$ the agglomeration time-scale is underestimated.
None the less the transformation still serves to indicate how
evolutionary time-scales
can be increased if initial cores of the same mass are moved to
larger radii with a correspondingly reduced disc mass.
\begin{figure*}
\centerline{\epsfig{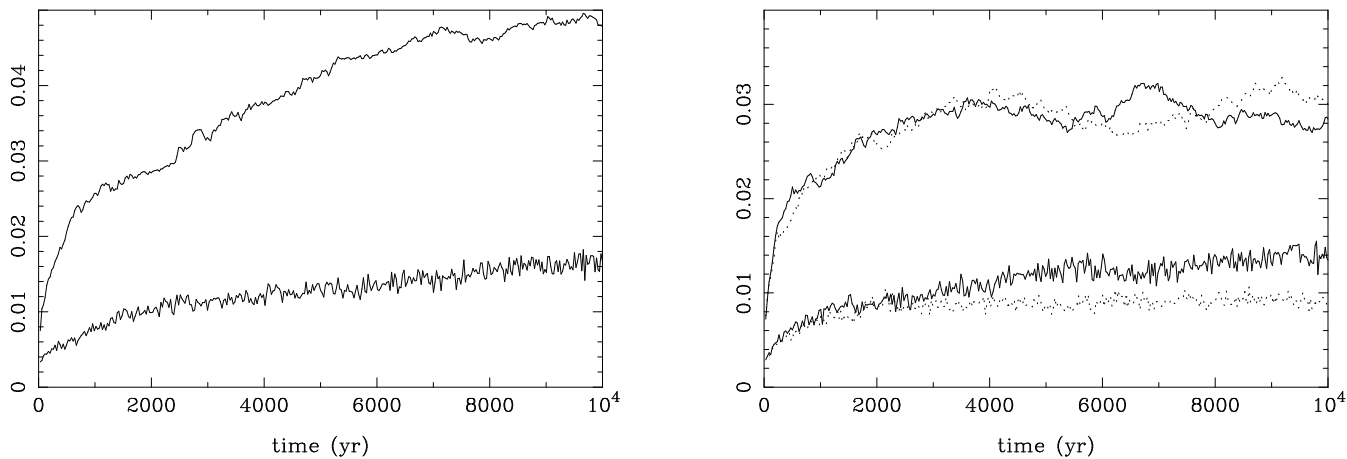}}
\caption{Time evolution of the mean eccentricity (uppermost curves)
and the mean value of $|z|/r$ (lowermost curves).
The left-hand panel is for a model without nebula torques included
in the force calculation. The right-hand panel shows data for
a run with eccentricity damping only (solid lines) and a run
with both eccentricity damping and inclination damping (dotted lines).}
\label{fig2}
\end{figure*} 
\begin{figure*}
\centerline{\epsfig{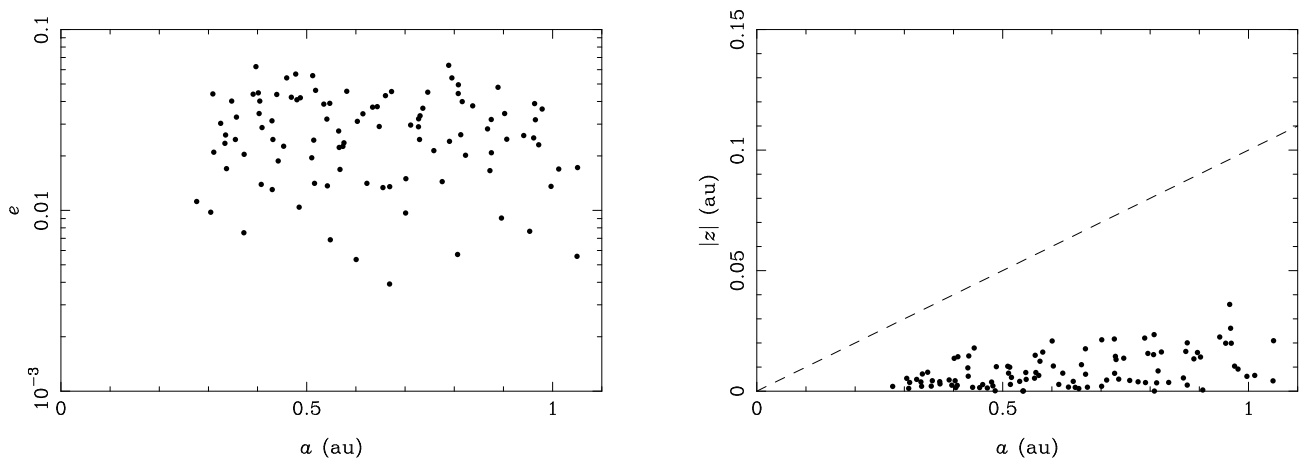}}
\caption{The eccentricity and vertical distributions of
particles in the model having $H/r=0.1$ and no migration or
inclination damping,
taken at a time of $10^4$\yr. In the right-hand panel
the dashed line refers to the scale height of the putative gas disc.}
\label{fig3}
\end{figure*} 

\subsection{Equilibrium eccentricity}

In Figure $2$ we compare the time evolution of the mean eccentricity
and the mean value of $|z|/r$, which is used to estimate
the characteristic value of $H_\alpha/r$. We do this
for models with and without damping, but for which we did not
include orbital migration or agglomeration. We recall that the
dynamical relaxation time-scale estimated above is
$t_R\sim 1400$\yr, at the mean radius. For the model that includes
gas with $H/r=0.1$ we find $t_e\sim 6800$\yr, at the mean radius
for $e\ll H/r$ from equation (\ref{efit}).

Without nebula interaction the model cores show eccentricity and vertical
thickness increasing monotonically with time.
The model that only uses eccentricity damping shows that a
quasi-equilibrium in the mean eccentricity is attained after
$\sim 4000$\yr. The average eccentricity is $\sim 0.3H/r$ and we
infer $H_{\alpha}/r\sim 0.3H/r$, broadly in agreement with our analytical
estimates outlined in the previous Section.

Comparison of the eccentricity-damped model with a similar model
that also includes inclination damping with $t_i=t_e$ shows that
the time-scale and level of the equilibrium eccentricity is
essentially the same as before.
In the former model, the equilibrium thickness is
established after a time $\sim 4000$\yr. In the latter
model, the equilibrium value is established after a time
$\sim 2000$\yr, at a level of about $40\%$ less than for the former model.
This value is minimal given the expectation of $t_i\geq t_e$.
These calculations assumed $M_{GD}=20M_J$, but can be compared
to other gas-disc masses by scaling $H \propto M_{GD}^{1/4}$; for example
the minimum mass Solar nebula with $M_{GD}=2M_J$ has the same equilibrium
as the model used here when $H/r_0 =0.06$.
Thus the aspect ratio of the protoplanet swarm relative to
the gas-disc aspect ratio increases as the gas-disc mass is reduced.

In Figure $3$ we give the eccentricity and vertical position
distributions at a time of $10^4$\yr, for the model without
inclination damping. The maximum eccentricity is $\sim 2/3H/r$.
The low level of eccentricity and inclination in all cases with damping puts
the protoplanet distribution well inside the putative disc gas
environment, as demonstrated in the right-hand panel of Figure $3$.
This situation promotes physical collisions
in a system of bodies with finite size, which we consider next.

\begin{figure*}
\centerline{\epsfig{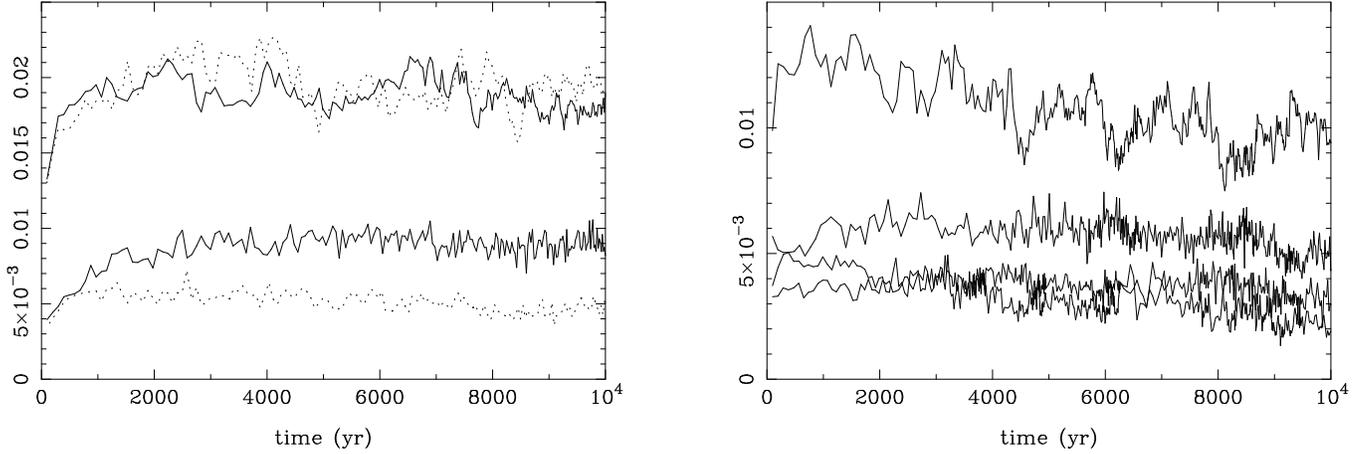}}
\caption{Time evolution of the mean eccentricity
and the mean value of $|z|/r$,
for models with damping, migration and agglomeration.
The left-hand panel shows data for the $H/r=0.07$ runs with
eccentricity damping only (solid lines) and with inclination
damping included (dotted lines). In both cases the mean eccentricity
has the higher value. The right-hand panel shows data for
the $H/r=0.05$ model (uppermost curves, for which the higher value
corresponds to the mean eccentricity), and also the $H/r=0.03$ model
(lowermost curves, for which the mean eccentricity initially has
the higher value).}
\label{fig4}
\end{figure*} 

\subsection{Core agglomeration models}

We present models with $H/r= 0.07$, $0.05$ and $0.03$, employing both
eccentricity damping and orbital migration. Again, $M_{GD} = 20M_J$.
For comparison we
repeated the model with $H/r=0.07$, but augmented with inclination damping
such that $t_i=t_e$. Additionally, there is one model without
damping or migration. For computational convenience the radius of
the central object is taken as $10R\sun$ in all cases with orbital
migration. First we consider the previously discussed equilibrium
in agglomerating simulations, and then protoplanetary growth
and orbital migration.
\begin{figure*}
\centerline{\epsfig{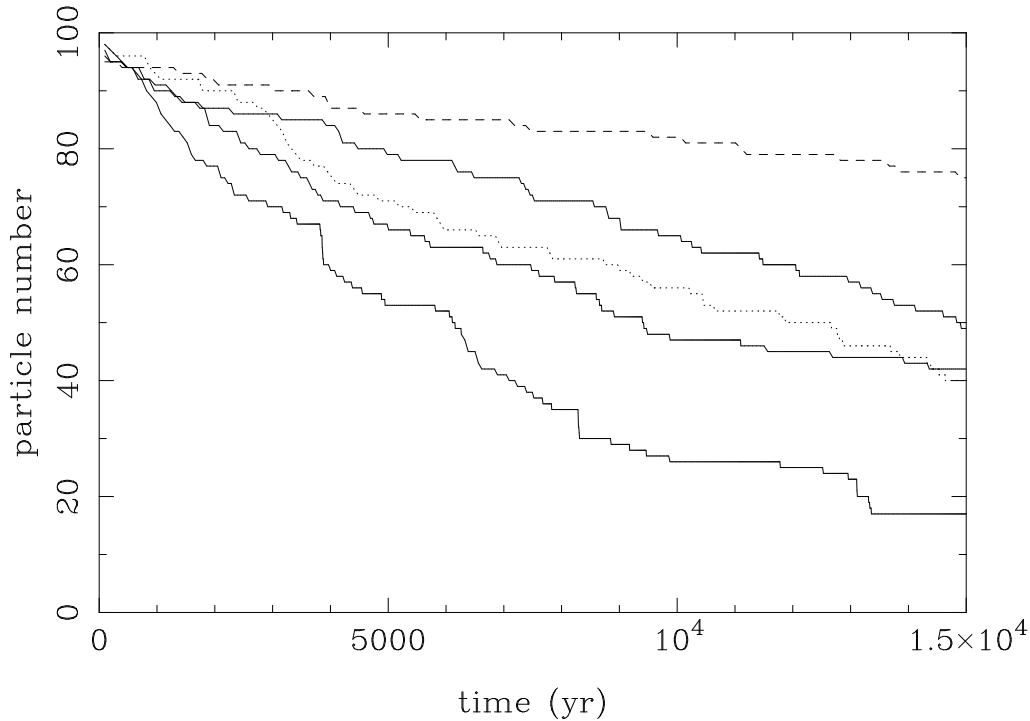}}
\caption{The total particle number in the simulations versus time.
The data for the model without nebula interaction is plotted with
a dashed line,
the data for the model with $H/r=0.07$ and inclination damping
included is plotted with a dotted line.
From highest to lowest final value the solid lines are for
models with eccentricity damping and orbital migration having
$H/r= 0.07, 0.05, 0.03$.}
\label{fig5}
\end{figure*} 

\subsubsection{Equilibrium}

Figure $4$ shows the average eccentricity and the mean
value of $|z|/r$ for all our models having migration and damping.
The runs with $H/r=0.07$ are qualitatively similar to the
analogous runs with $H/r=0.1$ of the previous Section, with a
value of $\sim0.3H/r$ for the average eccentricity and
$\sim0.2H/r$ for the semi-thickness to radius ratio, in the case without
inclination damping. Again,
inclination damping has little affect on the eccentricity
equilibrium, and a moderate affect on the particle-disc thickness,
being reduced by $\sim 40$\% in this case.
The model having intermediate gas-disc thickness $H/r=0.05$ gives
$\sim0.2H/r$ for the average eccentricity, and for
$H_{\alpha}/r$, but we note that the averages
show a long time-scale decline. This is due to the
diminishing particle number, which is a stronger effect in the thinner
gas-disc models than for the thicker gas-disc model (see below).
We note also that in this model the relative mean eccentricity has dropped
in comparison the previous cases, and it is about a half of the predicted
value. However, incorporation of migration and agglomeration
appears not to strongly affect the establishment of the predicted
velocity dispersion equilibrium in these cases.

The thin disc model with $H/r=0.03$ shows an equilibrium
semi-thickness to radius ratio $\sim 0.2H/r$, but the average eccentricity
is much smaller than for the thicker disc models at $\sim0.1H/r$,
although this has a value $\sim0.3H/r$ for the first $2000$\yr.
In this model the particle-disc thickness is comparable to the
mean Hill radius. Consequently the curvature of the Keplerian
orbits makes scatterings 
less efficient at pumping eccentricity than predicted and
so the average eccentricity seeks a lower equilibrium value.
Hence it is likely that the intermediate gas-disc thickness model is
close to being marginally applicable to our analysis, and the
agreement on the particle-disc thickness found in the thin gas-disc model
is coincidental since the initial mean Hill radius is $\sim 0.1H/r$.
In our models, in order to compare behaviour at small $H/r<0.05$
with the analytical predictions, we would require less massive planetesimals.
However, since our analytical results are relatively insensitive to
$m_\alpha$, similar qualitative behaviour would be expected.
\begin{figure*}
\centerline{\epsfig{file=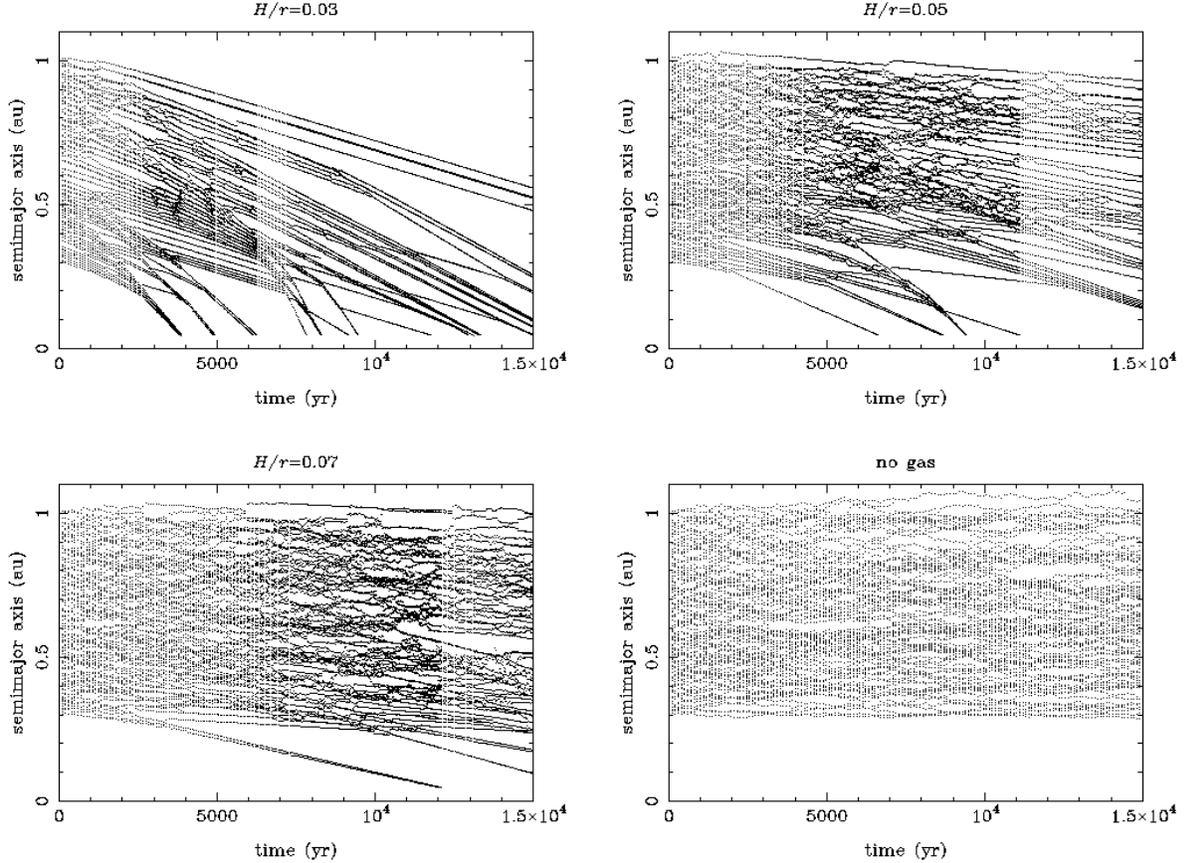,width=125mm,angle=270}}
\caption{Particle semimajor axis versus time for the models
without inclination damping, plus the model without nebula
torques. Note that the data is not uniformly spaced in time.}
\label{fig6}
\end{figure*} 

\subsubsection{Protoplanetary growth and orbital migration}

In Figure $5$ we show the total particle number versus time for
models with and without nebula interaction. The former calculations
include both damping and orbital migration. The agglomeration time-scale
for cores is reduced by including nebula interaction, owing to
vertical confinement of the particle distribution, limited radial
velocity dispersion and orbit crossing due to migration.
The rate of agglomeration for the model with $t_i=t_e$ and $H/r=0.07$
is of similar order to that with $H/r=0.05$ and $t_e$ only, being
only about $20$\% faster than that with $H/r=0.07$ and $t_e$ only. Hence
inclination damping does not very significantly enhance the agglomeration
time-scale provided $t_i\ga t_e$. In all cases with gas-disc models
included the accumulation rate is much faster than for the gas-free
model. We note that the model with $H/r=0.07$ has a migration rate
that is comparable to that expected for the minimum mass Solar nebula
with $M_{GD}=2M_J$ but having $H/r=0.02$.

In Figure $6$ we compare the particle trajectories for the models
with eccentricity damping and migration only and the case without
nebula interaction.
In all the cases with nebula interaction included we find
that several $\sim 0.5${--}$1M_\oplus$ cores build up and
undergo increasingly fast migration to the central regions such that
the time-scale to build up these masses is comarable to their
net migration time-scale. The system
thus appears to tend to the situation in which the remaining low
mass cores are sufficiently widely spaced that scatterings are no longer
effective in raising the eccentricities against damping.
However, as long as the core masses are not uniform,
differential migration of cores allows
further accumulation and rapid migration until uniformity is achieved,
or all matter is delivered to the central star.
If the former situation occurs, since $t_m $ increases linearly  with radius,
there can be no further mass accumulation, as the cores' orbital
separations do not decrease with time.

\section{Discussion}

We have derived eccentricity damping and orbital migration time-scales
for protoplanetary cores fully embedded in a model gaseous nebula disc
with $\Sigma \propto r^{-1.5}$. This has been done for $e$ significantly
larger than $H/r$, taking into account for the first time 
all effective contributing resonances $(n,m)$. We find that 
for $1>e\gg H/r$ circular orbit calculations are modified 
in a way indicated by the replacements:  
$t_e \rightarrow [e/(H/r)]^3t_e$ and $t_m \rightarrow -e/(H/r)t_m.$ 

This behaviour is in accord with the simple impulse model for tidal
interaction (Lin \& Papaloizou 1979). According
to this the coupling between the orbit
and disc would be expected to be
weaker for eccentric orbits owing to the larger relative velocity of
the protoplanet and local disc material.
The weaker gravitational interaction explains the generally longer
time-scales obtained in comparison to the circular orbit case.
We also find reversal of the direction of migration,
defined according to whether the orbital
angular momentum decreases or increases, is possible
for sufficiently eccentric orbits ($e\ga 1.1H/r$).
This can be thought of as a consequence of the
protoplanet spending most time interacting with the disc at apocentre,
where the relative velocity of the disc matter gives a predominating
contribution to net migration in the outward direction.

We note also that the circularization time-scale can be significantly
increased when the  orbit has  significant eccentricity.
Application of (\ref{efit}) indicates that for 
$M_{GD} =2M_J$ and $H/r=0.07$, a critical
mass core of $15M_{\oplus}$ has a circularization
time of $\sim 1.6\times10^3$\yr~at $10$\au.
If $e=0.35$ this is increased to $10^5$\yr, which is comparable
to the local nebula accretion time-scale.
Thus standard inward migration can be prevented if the eccentricity can be
maintained against decay by some process such as
gravitational interaction with either larger objects and/or
or a distribution of planetesimals with enough total mass. Accordingly gas
accretion onto a critical mass core in an  orbit with modest
eccentricity should be studied. In this context we note that the
notion of a clean annular gap being opened
in the disc by the action protoplanetary tides breaks down
for eccentric orbits because resonances giving positive and negative torques
exist at radii both less than and greater than the semimajor axis. In the
small eccentricity case the positive and negative torques are divided
separately between the opposing sides of the orbit (Ward 1997b).

We calculated the dynamical relaxation time-scale for a system of
cores and deduced the equilibrium eccentricity set up
when there is a balance between
pumping through gravitational scattering 
and damping through nebula interaction. We find
that the vertical distribution of the protoplanet swarm generally
remains well confined within the
gaseous envelope of the nebula. This supports our use of a two
dimensional analytical model for the tidal interaction.

We determined analytically, and confirmed empirically, that the
mean eccentricity and the thickness to radius ratio for the
protoplanet swarm are roughly $20${--}$30$\%
of the gas-disc aspect ratio, provided the characteristic size
of a protoplanet's Hill sphere is smaller than the latter. But we
note that there is a weak dependence on gas-disc and particle-disc mass, and
also radius. If the characteristic Hill radius is larger than or
comparable to the thickness to radius ratio then lower equilibrium
values of thickness and eccentricity occur. The vertical
confinement of the cores enhances the collision frequency and
consequently promotes more rapid agglomeration into larger objects.

\subsection{Protoplanetary growth}

With the aid of a numerical code we showed that the presence of a
gaseous disc can dramatically affect
the collisional evolution of a system of gravitating cores.
We found that several cores
increased in mass by a factor $\sim 10$ in only $\sim 10^4$\yr,
and underwent migration to the central star on the same time-scale,
for a gas-disc $\sim 10$ times more massive than the minumum
mass Solar nebula.
Our scaling for $t_m$ is such that the migration
time-scale for a protoplanet in a $M_{GD}=20M_J$ gas disc is
the same as for the same protoplanet in the $M_{GD}=2M_J$ minimum mass
Solar nebula gas disc, but with a moderately reduced $H/r$.
Additionally, the migration rate of our
adopted $0.1M_\oplus$ initial core isolation mass at $1$\au~scales to the
same value for a $1M_\oplus$ protoplanet at $10$\au.
Starting from an ensemble of equal mass cores embedded in the disc model
with $\Sigma \propto r^{-1.5}$, as described above, implies that rapid
gas accretion onto a massive core cannot occur before it plunges
into the star. Tanaka \& Ida (1999) reach similar
conclusions by considering the growth rate of a migrating $1M_\oplus$
protoplanet through a field of much smaller planetesimals.
By extrapolation of the growth rate, they find that for the
minimum mass Solar nebula, protoplanets could only build up to a
few earth masses before being consumed by the stellar envelope.

Thus the inward migration of at least one core must be halted short
of the stellar envelope
by some means at the initial stage of clearing, if a gas giant is to
be formed at all. If a halting mechanism, such as the
presence of a magnetospheric cavity, operates then
other cores would be delivered to the first one owing to orbital
migration. Providing the migration is not too rapid  the
inner body could feed on the inflowing
objects (Ward 1997a, Papaloizou \& Terquem 1999), presumably
giving a critical core in only $\sim 10^4$\yr. But we note
that no halting mechanism is known to apply to Jupiter at its present
location of $5$\au. Although it may be possible that the cores of
the outer planets were formed at much larger
disc radii with correspondingly longer migration time-scales,
and were possibly much closer together in the very early stages of
their formation, which could also allow for
the mutual excitation of their eccentricities and the consequent
stalling of inward orbital migration. It could also be
possible that planetesimal scatterings by the cores could help to
maintain their eccentricities (Hahn \& Malhotra 1999).

Vertical confinement of the protoplanet swarm through inclination
damping helps to reduce the agglomeration time-scale without increasing
the migration rate. However, for inclination damping time-scales
greater than the eccentricity damping time-scale this is
not a very significant effect. Even when the zero thickness
limit is approached, we still have the problem of type I orbital
migration causing very rapid inward migration of the most massive cores.

\section*{Acknowledgments}
This work was supported by PPARC grant GR/H/09454.

\label{lastpage} 
\end{document}